\documentclass[conference]{IEEEtran}
\IEEEoverridecommandlockouts
\usepackage{cite}

\usepackage{amsmath,amssymb,amsfonts}
\usepackage{algorithmic}
\usepackage{graphicx}
\usepackage{textcomp}
\usepackage{xcolor}
\def\BibTeX{{\rm B\kern-.05em{\sc i\kern-.025em b}\kern-.08em
    T\kern-.1667em\lower.7ex\hbox{E}\kern-.125emX}}
\begin{document}

\title{AI-Adaptive Virtual Reality for STEM: Engagement, Retention, and Learning Outcomes  \\
}

\author{\IEEEauthorblockN{Dr. Srinivasa K G}
\IEEEauthorblockA{
% \textit{Dept. of Data Science and Artificial Intelligence} \\
\textit{IIIT Naya Raipur}\\
India \\
srinivasa@iiitnr.edu.in}
\and
\IEEEauthorblockN{Akshit Rana}
\IEEEauthorblockA{
% \textit{Dept. of Computer Science and Engineering} \\
\textit{IIIT Naya Raipur}\\
India \\
akshit20100@iiitnr.edu.in}
\and
\IEEEauthorblockN{Aman Singh}
\IEEEauthorblockA{
% \textit{Dept. of Computer Science and Engineering} \\
\textit{IIIT Naya Raipur}\\
India \\
aman20100@iiitnr.edu.in}
\and
\IEEEauthorblockN{Kshitij Kumar Singh Chauhan}
\IEEEauthorblockA{
% \textit{Dept. of Computer Science and Engineering} \\
\textit{IIIT Naya Raipur}\\
India \\
kshitij20100@iiitnr.edu.in}
\and
}

\maketitle

\begin{abstract}
Immersive Virtual Reality (VR) presents a compelling opportunity to transform STEM pedagogy; however, empirical evidence evaluating AI-driven adaptivity, objective engagement metrics, and long-term knowledge retention remains sparse. This paper proposes and evaluates an AI-adaptive VR-based solar system learning module deployed on Meta Quest 2, wherein a machine learning model dynamically adjusts instructional difficulty based on real-time learner interaction logs including dwell time, gaze patterns, and response latency. A controlled between-subjects experiment was conducted involving 120 secondary school students (mean age $14.3$ years) across three schools in India, randomly assigned to three conditions: AI-adaptive VR, standard VR, and traditional classroom instruction. Engagement was measured objectively through eye-tracking fixation density and heart rate variability (HRV), alongside pre-test, immediate post-test, and delayed post-tests at one week and one month intervals. Statistical analyses encompassing paired $t$-tests, one-way ANOVA, and Structural Equation Modeling (SEM) reveal that the AI-adaptive VR cohort demonstrated significantly higher knowledge improvement scores (mean gain $= 38.4$ vs.\ $31.2$ vs.\ $10.7$), superior one-month retention ($p < 0.001$), and measurably elevated engagement across all biometric indicators. SEM confirmed immersion as a significant mediator between adaptive instruction and knowledge acquisition ($\beta = 0.63$, $p < 0.001$). Post-session self-directed learning exploration was recorded at $80\%$ among AI-adaptive VR participants versus $40\%$ in the traditional group. These findings establish an evidence-based framework for deploying AI-adaptive immersive learning environments in resource-varied STEM educational contexts.
\end{abstract}


\section{Introduction}
Virtual Reality (VR) technologies have witnessed significant integration across various fields, particularly in education, teaching, and training. Recent technological advancements have rendered VR more accessible and interactive, eliminating cumbersome cables and enhancing user freedom. The emergence of standalone VR Head Mounted Displays (HMDs) has further propelled the adoption of VR in educational contexts, promising improved engagement and immersive learning experiences. The use of virtual reality (VR) in education has been a growing area of interest, with a focus on its potential to enhance student experience, motivation, and learning outcomes \cite{p1,p2}. Immersive VR, in particular, has been identified as a promising tool for making learning more engaging and motivating \cite{p3}. However, the application of VR in higher education has been found to be more prevalent in science, engineering, and medical-related majors, with limited use in humanities and social sciences \cite{p4}. Despite these limitations, the overall impact of VR in education has been positive, with potential for further exploration and development.

\begin{figure}
    \centering
    \includegraphics[width=0.85\linewidth]{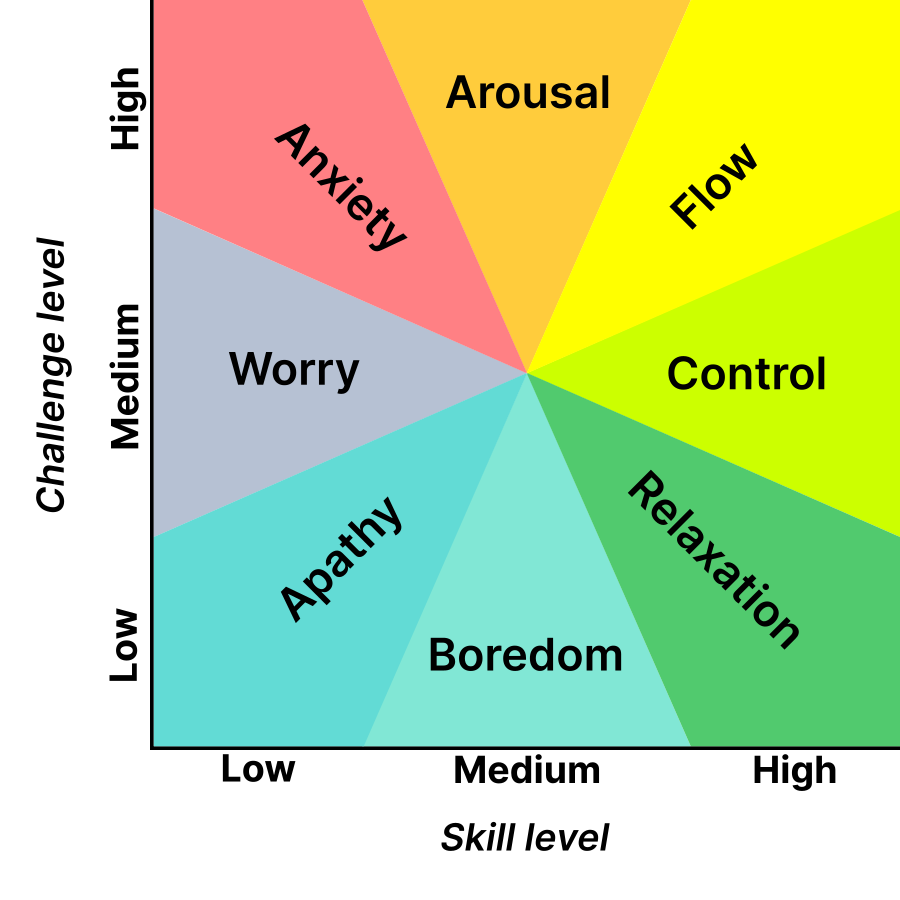}
    \caption{Flow mode: Challenges vs Skills \cite{p28}}
    \label{fig:flowmodel}
\end{figure}

Flow theory \cite{p28} as explained by Figure \ref{fig:flowmodel} asserts that the attainment of flow is contingent upon three essential conditions. Firstly, the activity should possess distinct objectives and a tangible sense of progress, thereby establishing a framework and direction. Secondly, the task necessitates the provision of prompt and unequivocal feedback, facilitating the negotiation of evolving demands and enabling the adjustment of performance to sustain the flow state. Lastly, a harmonious equilibrium between the perceived challenges inherent in the task and one's perceived skills is imperative. Confidence in the ability to successfully accomplish the task serves as a prerequisite for achieving the flow state. So, the learning activity must be endowed with well-defined objectives and visible progress, providing a structured and directional framework. Secondly, the learning task should offer immediate and clear feedback, allowing learners to adapt to changing demands and sustain the immersive learning experience. Additionally, there needs to be a delicate balance between the perceived challenges of the virtual reality (VR) learning experience and the learners' perceived skills, fostering confidence in their ability to navigate and succeed in the educational tasks presented in the VR environment.

VR exerts a substantial positive impact on creative design processes \cite{p16}, influencing critical stages such as design and planning, testing and revision, as well as fostering thoughts and appreciation. Moreover, VR significantly enhances the overall outcomes of creative design endeavors, contributing both high- and low-level effects on creative design and functional appropriateness, respectively. The learning process within the domain of VR and creative experiential learning undergoes a dynamic shift, progressing from concrete experiences to reflective observations, and then evolving into abstract conceptualization and innovative applications. Key behaviors, notably reflective observation and active engagement in innovative application, play pivotal roles in steering the trajectory of creative processes and shaping the resulting outcomes. As such, VR emerges not merely as a technological tool but as an immersive and transformative medium with the potential to redefine and elevate educational experiences.

This study revolves around the core objective of addressing the intrinsic challenges associated with student engagement and motivation in the domain of science education, focusing specifically on the solar system. Traditionally, science topics, including astronomy, are often disseminated in a theoretical manner due to various constraints, leading to a potential lack of practical and immersive interaction in foundational topics such as the solar system. This work undertakes a comparative analysis of two distinct teaching methodologies: conventional instruction and an immersive virtual reality (VR) experience. The emphasis lies in adapting these methodologies to both confined and non-confined learning scenarios, ensuring contextual relevance amid the transformative shift in educational practices instigated by the ongoing pandemic.

Primary goal of this study is bifurcated. It aims to contribute nuanced insights into the efficacy of VR in enhancing pivotal factors such as class engagement, learning motivation, and knowledge acquisition among students in the context of the solar system. Additionally, it acknowledges the pressing need for practical and engaging methodologies, especially given the dynamic nature of contemporary educational landscapes. This paper distinguishes itself by providing a robust foundation for statistically evaluating the impact of VR on key educational metrics. Furthermore, it seeks to delineate the unique contributions of VR in science education, offering an immersive approach that transcends traditional teaching methods. The research rigorously analyzes the comparative effectiveness of VR and conventional approaches in order to contribute substantiated findings to the discourse on enhancing science education methodologies.

\section{Background}
\subsection{VR in education}
As VR transforms entertainment experiences, its application in education has the potential to revolutionize traditional methods of knowledge acquisition both within and outside school classrooms. Educators developing VR tools need a thorough understanding of building and deploying educational programs that align with the technology's principles and cater to learners' needs, preparing them for the demands of the 21st century. Researches have demonstrated that executive function is a vital component in learning from VR \cite{p26}. Therefore, instructors should think about the amounts of distraction that may be present in an VR lesson prior to including this as an instructional method to a class. In the context of this research, the exploration of VR's impact on learning experiences and understanding how VR influences engagement, motivation, and knowledge acquisition of the students is a critical aspect of advancing educational technology and pedagogy. This study seeks to contribute to the existing body of knowledge by rigorously examining the effectiveness of a VR-based Solar System Experience on high school students' understanding and learning outcomes. Research findings highlight that students who underwent preliminary instruction from the teacher exhibited significantly superior performance compared to their counterparts \cite{p25}. This underscores the crucial necessity of furnishing learners with foundational information and concepts before engaging them in an Immersive Virtual Reality learning experience.

\subsection{VR's Impact on Spatial Memory}

VR's adaptability allows for personalized and adaptive learning experiences. Learners can be exposed to spatial challenges tailored to their proficiency levels, ensuring that the complexity of tasks aligns with individual cognitive capacities. VR technology, beyond its role in enhancing engagement and motivation, extends its influence to various facets of education. One critical area is spatial memory, where VR immerses learners in virtual environments, allowing freedom of movement and interaction. \cite{p9} and \cite{p10} investigated the impact of VR on spatial learning and Memory of Loci, revealing improved retention and recall. The immersive nature of VR, exemplified by high-end systems like Oculus Rift, enables users to view scenes from multiple angles, potentially enhancing memory formation. While VR offers promising prospects for spatial memory enhancement, challenges exist. The cognitive demand imposed by immersive VR experiences can lead to extraneous cognitive processing, affecting learning performance \cite{p11}. Additionally, \cite{p12} observed increased processing demands on working memory, resulting in decreased knowledge acquisition. The intricate balance between cognitive load and the benefits of VR necessitates careful instructional design to optimize learning outcomes. , VR's impact on spatial memory extends beyond traditional educational approaches, offering immersive and dynamic learning environments conducive to spatial skill development. The ability to enhance cognitive mapping, foster navigational skills, facilitate transferability to real-world contexts, and provide adaptive challenges positions VR as a valuable tool for spatial memory improvement in educational settings.

\subsection{VR's Role in Empathy Training}

Moving beyond spatial memory, VR's influence extends to empathy training, providing users with immersive experiences that foster understanding. Various applications and and experiences have been designed for empathy-building. These highlight VR's potential to facilitate perspective-taking tasks. The immersive capabilities of VR contribute to lasting learning experiences, particularly in understanding diverse perspectives. Despite the potential benefits, challenges in VR applications for empathy training exist. The degree of immersion depends on hardware and software quality, with suboptimal experiences detracting from the intended impact \cite{p13}. Issues such as visual aberrations, discomfort from extended use, and limitations in maintaining immersion can diminish the effectiveness of VR in empathy training. The Virtual Human Interaction Lab at Stanford University created "1000 Cut Journey," \cite{p14}; an immersive VR experience that enables users to walk in the shoes of a Black male named Michael Sterling, experiencing racism at different life stages. This type of embodied simulation contributes to a more profound understanding of the impact of systemic issues and biases. Similarly, in \cite{p15} authors conducted a study using VR to help teacher trainees experience being both a student and a teacher. The results indicated a significant improvement in empathy related to understanding the experiences of racially and ethnically different students. VR's immersive perspective-taking tasks contribute to lasting changes in attitude and perception. Empathy training in VR goes beyond passive observation; it actively engages users in perspective-taking exercises. The ability to see the world through someone else's eyes enhances cognitive and affective empathy, encouraging users to consider alternative viewpoints and fostering a more inclusive mindset.

\subsection{VR in Distance Learning}

Beyond empathy, VR has the potential to reshape distance learning by providing immersive and collaborative environments, enhancing language learning experiences, and addressing the limitations of traditional online platforms. \cite{p17} discuss the potential of multi-user virtual campuses for distance education, emphasizing the capacity of VR to connect learners across distances. However, challenges include the rapidly changing software landscape and the need for effective content creation for a small user base. The COVID-19 pandemic has heightened interest in distance education, making VR a potential solution. \cite{p18} found VR useful for task-based language distance learning, emphasizing the importance of correct utilization for immersive learning methods. By leveraging the immersive capabilities of VR, educational institutions can overcome some of the challenges associated with traditional online learning methods, providing students with interactive and engaging environments that mimic real-world scenarios. Despite the potential benefits, integrating VR into distance learning poses challenges that educators and researchers must address. Issues such as the rapidly changing software landscape, the need for a VR-specific pedagogy, and considerations around gender-related aspects when implementing VR in the classroom demand careful attention. 

\subsection{Pedagogical Considerations in VR Implementation}

 Recognizing that VR is a transformative technology, educators and researchers must adapt pedagogical approaches to leverage its unique affordances effectively. In implementing VR for education, lack of VR-specific pedagogy poses a significant challenge. While VR can be integrated into existing paradigms, a dedicated pedagogical framework is essential \cite{p19}. The absence of informed pedagogy may contribute to mixed findings in literature, necessitating the development of instructional designs aligned with models like constructivism. Pedagogical considerations are critical in leveraging VR effectively. Constructivism emerges as a foundation, emphasizing meaningful experiences that align with VR's immersive nature. Recognizing the need for VR-specific pedagogy, recent frameworks like Southgate's Actioned Pedagogy \cite{p20} for Immersive Learning provide guidance for educators. However, ongoing research is essential to refine instructional designs that capitalize on VR's unique affordances. This approach encourages educators to consider key aspects related to the teacher, learners, and the technology itself. Three Key Elements of Actioned Pedagogy are as listed.

\begin{itemize}
        \item \textbf{Teacher's Role:} In VR environments, educators become facilitators guiding students through immersive experiences. Understanding how to manage and guide students in a VR setting is crucial for effective learning outcomes.
        \item \textbf{Learner-Centered Approaches:} Tailoring VR experiences to meet individual learning needs enhances engagement. Allowing students to explore and make choices within the VR space promotes autonomy and self-directed learning.
        \item \textbf{Technology Integration:} Ensuring that the chosen VR technology aligns with pedagogical goals is paramount. Compatibility, ease of use, and the ability to facilitate meaningful learning experiences should drive technology selection.
\end{itemize}

Frameworks, such as the one developed by Lege, Bonner, Frazier, and Pascucci \cite{p21}, offer a structured approach to analyzing commercial off-the-shelf (COTS) VR applications. This involves assessing the applicability of VR tools in the classroom, considering factors like content relevance, interactivity, and alignment with educational objectives. Some framework components are as following.

\begin{itemize}
        \item \textbf{Content Relevance:} Evaluating how well VR content aligns with curriculum goals and enhances the learning experience.
        \item \textbf{Interactivity:} Assessing the level of interactivity within VR applications to ensure active student engagement.
        \item \textbf{Alignment with Objectives:} Ensuring that VR tools support specific educational objectives and contribute to the overall learning outcomes.
\end{itemize}

While leveraging existing pedagogical frameworks, there is a recognized need for a VR-specific pedagogy \cite{p22}. This involves understanding how to design and deploy educational programs uniquely suited to VR technology. Following are key considerations:

\begin{itemize}
        \item \textbf{Immersive Learning Experiences:} Prioritizing the creation of immersive, authentic learning experiences that capitalize on VR's capacity for simulation.
        \item \textbf{Learning Analytics:} Incorporating tools for gathering and analyzing data from students' interactions within VR environments to inform instructional decisions.
        \item \textbf{Assessment Strategies:} Designing assessment methods that align with the interactive and experiential nature of VR learning, moving beyond traditional testing approaches.
\end{itemize}

\subsection{Cognitive Demand and Immersion Breaking}

Given the finite cognitive resources of the human brain, educators must employ scaffolding techniques to support learners in managing cognitive load during VR experiences. The complex nature of immersive VR may create challenges in managing extraneous cognitive processing, necessitating thoughtful design to ensure learners focus on essential material.Despite the potential, cognitive demand remains a challenge in VR education. Issues such as cognitive load, immersive breaking, and unfamiliarity with VR necessitate strategic planning. \cite{p23} discusses introduction of summary tasks between VR sessions mitigated cognitive load without compromising engagement, highlighting the importance of careful consideration in designing VR educational experiences. As VR gains popularity, understanding and addressing challenges become imperative. In \cite{p20} author emphasizes measured introduction to VR in classrooms, allowing students time to familiarize themselves. Striking a balance that maximizes engagement and motivation while avoiding cognitive overload is crucial for achieving desired learning outcomes.

\section{Literature Survey}
The literature review encompasses several studies examining the role of Virtual Reality (VR) in educational contexts. In a review focused on the school setting, immersive content is recognized as a playful alternative to traditional lectures. However, the study underscores the limited research on potential risks, especially for young children, and advocates for a cautious introduction of immersive applications in schools, emphasizing the need for ongoing investigative assessments \cite{p1}.

Exploring VR in higher education, positive effects on students' behaviors, cognition, and learning outcomes are reported. Notably, there is a concentration of applications in science, engineering, and medical-related majors, revealing a gap in humanities and social science disciplines. Challenges identified include adverse student reactions, network conditions, and the imperative for extensive development and testing of VR educational resources \cite{p2}.

A systematic literature review on the use of VR as a pedagogical tool highlights its application in adult occupational training and creative design practices. Challenges include restricted implementation in educational settings, unresolved technical and social issues, and limited student experience in virtual environments \cite{p5}.

Emphasizing the importance of a rigorous methodological approach, another study notes the significant advantage of immersive virtual reality (I-VR) in education, particularly in science and engineering. Challenges identified encompass a relatively homogeneous application, scarcity of longitudinal studies, concerns about detrimental effects, and the demand for robust theoretical and experimental evidence \cite{p6}.

Addressing the implications of VR in arts education, a research analysis identifies driving agents and exponential growth in related articles. However, limitations are acknowledged in the intrinsic nature of quantitative bibliometric analysis, suggesting a need for methodological extensions in future research \cite{p7}.

In a comprehensive review exploring the promise and challenges of VR in education, the study highlights the difficulty faced by the educational field in keeping pace with rapid technological evolution. Challenges include a scarcity of research, a lack of comprehensive understanding regarding potentials and challenges, and recommendations lacking a foundation in a thorough literature review \cite{p8}.

In light of these studies as summarized in Table \ref{tab:literature_summary}, our research focuses on designing a sample VR game for educational purposes and analyzing its impact on students using statistical methodologies. The insights gathered from these literature reviews guide our approach, helping us anticipate potential challenges and emphasizing the importance of a meticulous design and assessment process.

\begin{table*}[htbp]
  \centering
  \begin{tabular}{p{2.5cm}p{3.5cm}p{5cm}p{5cm}}
    \hline
    \textbf{Title} & \textbf{Abstract Summary} & \textbf{Observations} & \textbf{Challenges and Limitations} \\
    \hline
    VR in school context \cite{p1} & Immersive content as playful alternatives. & Assessing VR and AR in education, articulating guidelines. & Limited research on risks, necessitating additional studies and gradual introduction in schools. \\
    VR in higher education \cite{p2} & Positive effects on students, predominantly in specific majors. & Limited diversity in applications, challenges include adverse reactions and network conditions. & Lack of examination across diverse majors, time-consuming development. \\
    VR as Pedagogical Tool \cite{p5} & Encourages involvement in adult occupational training. & Limited implementation, unresolved technical and social issues. & Restricted application, technical and social challenges. \\
    Immersive VR in education \cite{p6} & Rigorous methodology essential, significant advantages reported. & Homogeneous application, scarcity of longitudinal studies, concerns about detrimental effects. & Limited application diversity, need for rigorous methodology. \\
    VR in Arts Education \cite{p7} & New directions for research, exponential growth. & Identifies primary driving agents, acknowledges limitations. & Bibliometric analysis limitations, recommends methodological extensions. \\
    VR in education: Promise and Challenge \cite{p8} & Advancing VR practicality in education. & Challenges in keeping abreast of technological evolution. & Scarcity of research, deficiency in understanding potentials and challenges. \\
    \hline \\
  \end{tabular}
  \caption{Summary of Literature Survey}
  \label{tab:literature_summary}
\end{table*}

\section{Designing the Virtual Solar System Experience}
The development of the Virtual Solar System Experience (VSSE) involved a blend of game development tools and immersive virtual reality (VR) hardware. In this section, we delve into the technical details of hardware and game development engine used for the purpose if creating a sample virtual solar experience.
\subsection{Unity as the Game Development Framework}

Unity served as the foundational framework for the Virtual Solar System Experience (VSSE), providing an extensive suite of tools and features conducive to the creation of a dynamic and interactive three-dimensional (3D) environment.

\subsubsection{High-Fidelity Spatial Representation}
Leveraging Unity's spatial computing capabilities, the VSSE achieved a precise and realistic representation of the solar system. Unity's extensible architecture facilitated the integration of complex spatial data, ensuring accurate planetary positions, sizes, and movements within the VR environment.

\subsubsection{Realistic Visual Rendering}
The Unity Asset Store, a repository of 3D models and visual effects, was instrumental in realizing realistic environments. High-quality assets were seamlessly integrated to depict planetary surfaces, atmospheric phenomena, and dynamic lighting, contributing to an immersive visual experience.

\subsubsection{Interactive Scripting and User Engagement}
Unity's scripting functionalities were harnessed to introduce interactive elements into the VSSE. Users, equipped with Meta Quest 2 controllers, could engage in intuitive interactions, manipulating virtual surroundings, zooming in on planets, and observing dynamic planetary rotations. The incorporation of Unity's physics engine facilitated the realistic portrayal of gravitational forces, enhancing the educational value of the VR experience.

\subsection{Integration with Meta Quest 2}
\subsubsection{Technical specifications}
\begin{figure}
    \centering
    \includegraphics[width=0.75\linewidth]{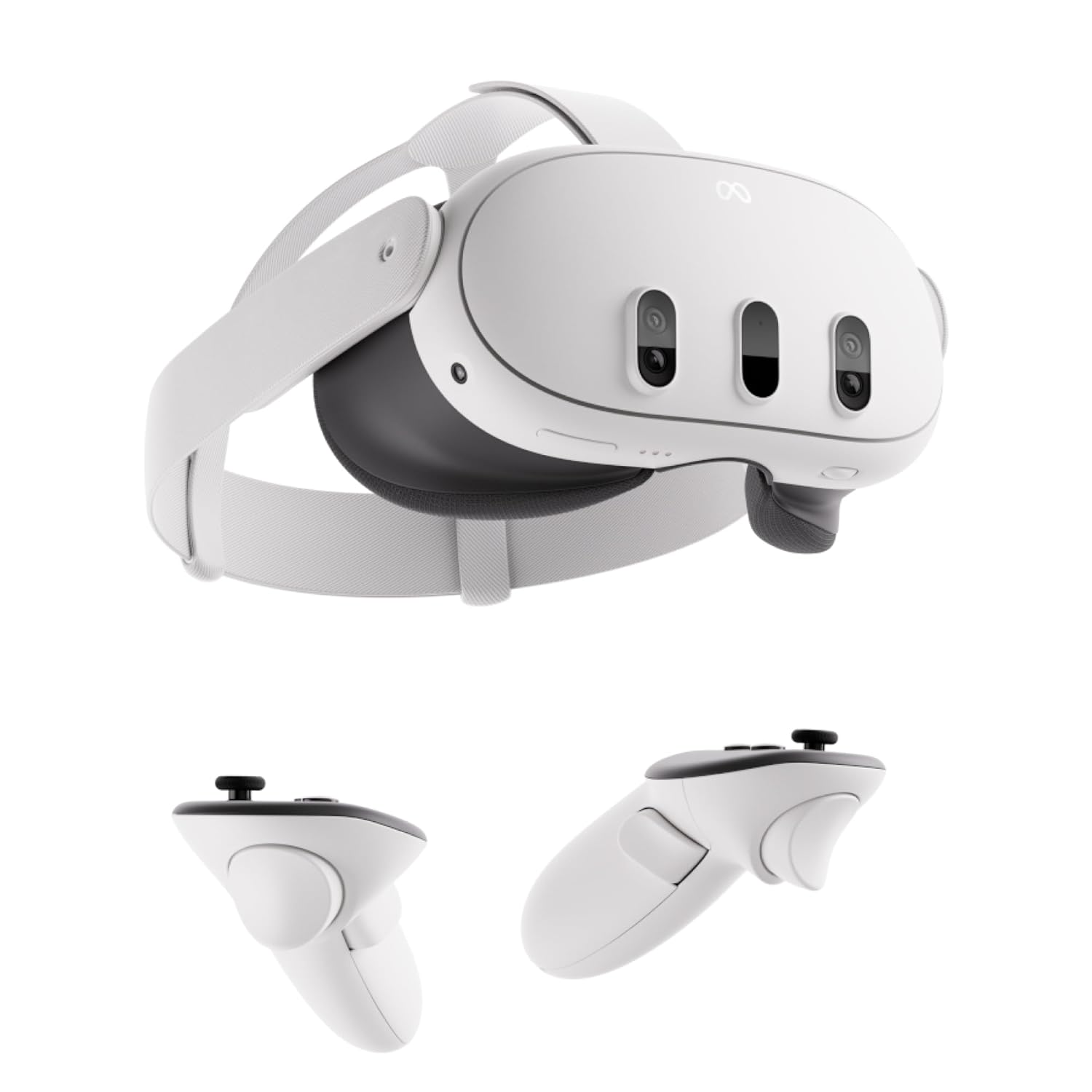}
    \caption{Meta Quest 2: VR Headset use in this research}
    \label{fig:hardware}
\end{figure}

The Meta Quest as depicted in Figure \ref{fig:hardware}, offers an immersive and accessible platform featuring a fast-switch LCD display with a resolution of 3664 x 1920 pixels, powered by the Qualcomm Snapdragon XR2 processor. The headset provides a visually appealing and technologically advanced foundation. With 6GB of RAM, six degrees of freedom (6DOF) tracking through Oculus Insight, and Oculus Touch controllers, the Meta Quest 2 ensures a responsive and interactive learning environment. The integrated speakers and along with compatibility with a wide range of VR applications on the Oculus platform, contribute to a comprehensive and engaging educational experience. These specifications prove particularly relevant to the development of a Virtual Solar System, as the high-resolution display and powerful processor facilitate the rendering of detailed celestial bodies, while 6DOF tracking and interactive controllers, such as the Oculus Touch, enable students to navigate and interact seamlessly within the virtual entities. The immersive audio and expansive storage options enhance the overall educational impact, making the Meta Quest 2 an ideal hardware solution for fostering an in-depth understanding of astronomy among students. 

\subsubsection{Hardware Compatibility and Optimization}
The Meta Quest 2 VR headset, chosen for its wireless capabilities and immersive features, was seamlessly integrated with Unity to optimize the Virtual Solar System Experience. Unity's compatibility with Meta Quest 2 facilitated streamlined deployment, ensuring optimal utilization of the VR headset's hardware capabilities. Rigorous optimizations were implemented to address hardware constraints, delivering a smooth and visually immersive VR experience without compromising performance.

\subsubsection{User Interaction in Virtual Reality}
Meta Quest 2's controllers, with Unity's input system, enabled nuanced user interactions within the VR space. The mapping of controller movements and gestures to specific actions allowed users to navigate the solar system, select celestial bodies, and access detailed information seamlessly.

\begin{figure*}
    \centering
    \includegraphics[width=0.95\linewidth]{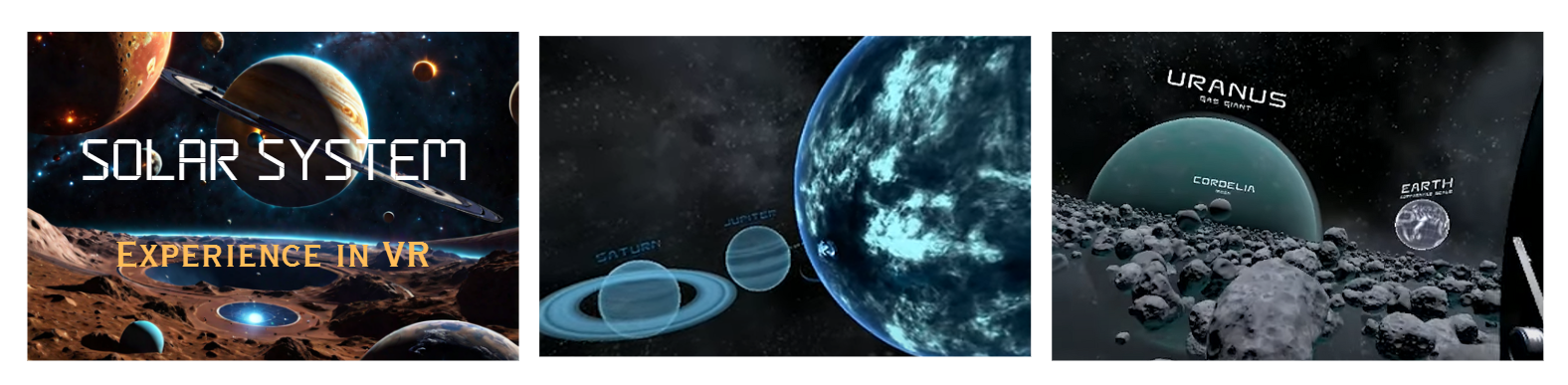}
    \caption{Sample overview of virtual solar system experience}
    \label{fig:game}
\end{figure*}

\subsection{Overview of the Virtual Solar System Experience}
Virtual solar system experience let students embark on an educational journey within a VR environment that accurately depicts the solar system's spatial dynamics and key information. Figure \ref{fig:game} presents a sample overview of the experience. Unity's scripting functionalities enrich this experience by introducing interactive elements, allowing users to interact with virtual surroundings and access detailed information about planets, including temperatures and atmospheric compositions. The capability to interact across six degrees of freedom (6DOF) allows for seamless observation of planetary rotations, revolutions and other similar concepts.

\section{Methodology}
The study employed a quasi-experimental design with a total of 30 participants, divided into two groups: the experimental group (n=15) and the control group (n=15). Due to the limited sample size, the results should be interpreted with caution and considered as preliminary findings. Future studies should aim for larger and more diverse samples to validate these results.

\subsection{Theoretical Framework}
The VR learning experience was designed based on Mayer's Cognitive Theory of Multimedia Learning (CTML), which posits that learning is more effective when information is presented in both visual and verbal forms. This theoretical framework guided the development of the VR content to ensure it effectively facilitated learning.

\subsection{Instructional Design Model}
The instructional design of the VR learning tool followed the ADDIE model (Analysis, Design, Development, Implementation, Evaluation). This systematic approach ensured that the learning experience was well-structured and pedagogically sound.

\subsection{Unique Features of the VR Tool}
The VR platform incorporated several unique features to enhance learning:
\begin{itemize}
    \item \textbf{Interactive Simulations}: Students could interact with the solar system model, manipulate celestial bodies, and observe real-time changes.
    \item \textbf{Real-time Feedback}: Immediate feedback was provided to reinforce correct actions and guide learners through incorrect ones.
    \item \textbf{Adaptive Learning Paths}: The tool adjusted the difficulty level based on the learner's performance, providing a personalized learning experience.
\end{itemize}
This section delineates the methodological framework employed in our research, which seeks to design, develop, and evaluate a Virtual Reality (VR) based educational game tailored for high school students in the Indian educational context. Grounded in the principles of Design Science Research (DSR) \cite{p29}, our methodological trajectory encompasses an intricate interplay of stakeholder engagement, data analysis, game design informed by educational requirements, and a statistical evaluation process.

Central to DSR is the development of artifacts that not only showcase technological prowess but also align with educational objectives. The primary predictor of learners' reflective thinking and comprehension in an educational game-based immersive virtual reality application is the clarity of goals \cite{p30}. In the case of this study, the virtual solar system experience was designed to offer an engaging exploration of the solar system, allowing students to interact with planets, observe their motion, spin, relative speeds of rotation and revolution, sizes, and access additional information such as temperatures through immersive VR experiences. Each interactive element was consciously crafted to address specific learning outcomes outlined in the Indian high school science curriculum.

\subsection{Participants}
Research design aimed to systematically investigate the impact of the VSSE on learning outcomes, ensuring a comprehensive evaluation of its efficacy in enhancing students' understanding of the solar system. The study engaged a cohort of high school students from a secondary educational institution in Lucknow, India. A total of 30 students, with an average age of 14, participated in the research. The selection aimed to ensure representation across various grades and academic levels to capture diverse perspectives and potential variations in prior knowledge of astronomical concepts. Random assignment was employed to allocate participants to either the Virtual Solar System Experience (VSSE) group or the traditional learning group.

\subsection{Design}

A controlled experimental design was implemented to systematically assess the impact of the Virtual Solar System Experience (VSSE) on high school students' comprehension of solar system concepts. The study employed a between-subjects design, with participants distributed across three distinct sets, each exposed to different learning methodologies. Students were randomly assigned to two sets: Set-1 and set 2, each containing 15 participants. Random assignment ensured the equivalence of groups, minimizing potential individual differences. This strengthened internal validity, allowing for more reliable inferences.

\subsection{Conditions}

\begin{enumerate}
    \item \textbf{Set-1 (VR-Based Experience):} Students engaged with the solar system concepts through the VR-based experience.
    
    \item \textbf{Set-2 (Blinded Traditional Method):} Participants experienced traditional teaching methods without knowledge of their involvement in an experiment, mitigating potential observation biases.
    
\end{enumerate}

% Section: Procedure
\subsection{Procedure}

\begin{enumerate}
    \item \textbf{Pre-assessment:} Participants underwent a pre-assessment to establish baseline understanding of solar system concepts.
    
    \item \textbf{Learning Phase:}
    \begin{itemize}
        \item Set-1: Engaged in the VR-based experience.
        \item Set-2: Unaware of the experiment, experienced traditional teaching.
    \end{itemize}
    
    \item \textbf{Post-assessment:} Following the learning phase, participants completed a post-assessment measuring changes in understanding.

    \item \textbf{Feedback:} Participants provided qualitative insights into their experiences with both methodologies.
\end{enumerate}

% Section: Data Analysis
\section*{Data Analysis}

Quantitative data from pre and post-assessments underwent statistical analysis. Qualitative feedback from Set-3 informed the interpretation of results. This approach allowed for a robust examination of the impact of the VSSE on participants' understanding while controlling for initial differences. The combined utilization of MANCOVA, ANOVA, descriptive statistics, and correlation analysis forms a comprehensive statistical framework for the study. MANCOVA serves as the primary analytical tool, allowing a focused examination of the specific impact of the VSSE while controlling for initial differences. ANOVA complements this by providing a broader perspective on group differences, offering insights into variations in learning outcomes among different teaching methodologies. Descriptive statistics contribute a succinct and informative summary of the central tendencies and variabilities within each group, enriching the overall interpretation of results. Furthermore, correlation analysis enhances understanding of the intricate relationships between engagement and learning outcomes, providing a nuanced perspective on the dynamics of active participation in the VR-based learning environment.  

\subsection{Descriptive Statistics}

The utilization of descriptive statistics, including measures such as mean scores and standard deviations, serves as a crucial aspect of the data analysis strategy. Descriptive statistics offer a concise and informative summary of the central tendency and variability within the dataset.

In this context, mean scores provide an overview of the average performance of participants within each set, indicating the typical level of understanding. Simultaneously, standard deviations elucidate the spread or variability around the mean within each group. This information is invaluable for gauging the consistency and reliability of the learning outcomes observed in different teaching conditions.

\subsection{Analysis of Variance (ANOVA)}

The exploration of group differences was facilitated by the application of Analysis of Variance (ANOVA). ANOVA is a powerful statistical tool that assesses variations in means among multiple groups. In the context of the study, ANOVA could be effectively employed to compare the post-assessment scores among the three distinct sets: Set-1 (VR-Based Experience), Set-2 (Blinded Traditional Method), and Set-3 (Comparative Feedback).

By incorporating ANOVA into the analytical repertoire, valuable insights into the overall differences in learning outcomes among participants exposed to different teaching methodologies can be gained. This complements the more focused examination provided by MANCOVA.

\subsection{Correlation Analysis}

A detailed exploration of potential relationships between variables is facilitated through correlation analysis. In the context of the study, correlation analysis could be applied to examine the association between the time spent in the VR-based experience and post-assessment scores. By quantifying both the strength and direction of this relationship, a nuanced understanding emerges regarding the degree to which active involvement in the VR-based educational milieu correlates with an augmented comprehension of solar system concepts.

Correlation analysis is particularly relevant for understanding the nuanced dynamics between engagement with the VR-based experience and subsequent learning outcomes. By quantifying the strength and direction of this relationship, a discerning perspective on the extent to which active participation in the VR-based learning environment correlates with enhanced understanding of solar system concepts can be gained.

By employing the diverse set of statistical methods as described in Table \ref{tab:desc_exp}, the analysis was able to to derive robust and reliable conclusions.

\subsection{T-Test Statistics Analysis}

In-depth scrutiny of the statistical differences between two sets of data is facilitated through the application of t-test statistics. Within the purview of this research, t-test analysis serves as a pivotal tool to assess the significance of disparities in learning outcomes between participants who engaged in the VR-based experience and those subjected to traditional teaching methods. T-test statistics assume paramount importance in gauging the statistical significance of observed variations in academic performance between the two groups. By evaluating the mean differences and standard deviations, this statistical method discerns whether the discrepancies in post-assessment scores are merely due to random chance or if they are indicative of a substantive effect stemming from the distinct pedagogical approaches.

This analytical approach provides a quantitative framework for determining the practical relevance of the observed differences in learning outcomes. The t-statistics and associated p-values derived from this analysis offer a robust foundation for interpreting the substantive impact of the VR-based learning experience. In essence, t-test statistics unravel the extent to which the VR-based educational intervention contributes to measurable improvements in academic achievement compared to conventional teaching methods.

\begin{table*}[htbp]
\centering
\begin{tabular}{p{2.5cm} p{5cm} p{3.5cm} p{3.5cm} p{3.5cm}}
\hline
\textbf{Set} & \textbf{Conditions} & \textbf{Statistical Methods} & \textbf{Remark} \\
\hline
\textbf{Set-1} & 
Participants experienced VR-Based Experience. & 
t-Test Statistics &
Assess the specific impact of VR-Based Experience on learning outcomes, controlling for initial differences. \\
\hline
\textbf{Set-2} & 
Participants experienced Blinded Traditional Method without awareness of being in an experiment. & 
ANOVA (Analysis of Variance), Descriptive Statistics &
Examine overall differences in learning outcomes among groups unaware of the experimental nature, using ANOVA. Utilize descriptive statistics for an overview. \\
\hline
\textbf{Both Set 1 and Set 2} & 
Post experimental survey on usability, engagement and similar aspects & 
Qualitative-feedback, Descriptive-Statistics, Correlation Analysis &
Provide a direct comparative feedback on the effectiveness of both methodologies. descriptive statistics for overview, and correlation analysis for exploring relationships. \\
\hline \\
\end{tabular}
\caption{Overview of Experiment}
\label{tab:desc_exp}
\end{table*}

\section{Results}
\subsection{Descriptive analysis}
\begin{figure}
    \centering
    \includegraphics[width=0.95\linewidth]{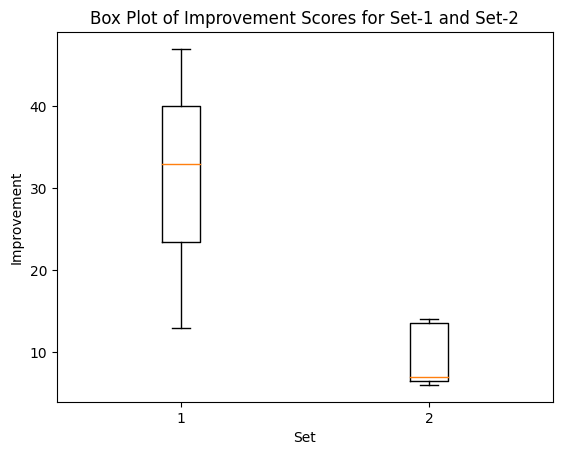}
    \caption{Analysis of knowledge improvement scores}
    \label{fig:knowledge_imp}
\end{figure}

This section presents a descriptive analysis of the data collected from the Virtual Solar System Experience (VSSE) research study. The analysis encompasses participants' engagement, class experience ratings, and knowledge improvement scores for both Set-1 and Set-2. Table \ref{tab:descanalysis} summarizes the engagement levels, average class experience ratings, and average knowledge improvement scores for participants in Set-1 (VR-Based) and Set-2 (Traditional). Knowledge improvement scores have been visualised using box-plot in Figure \ref{fig:knowledge_imp}.

\begin{figure*}
    \centering
    \includegraphics[width=0.95\linewidth]{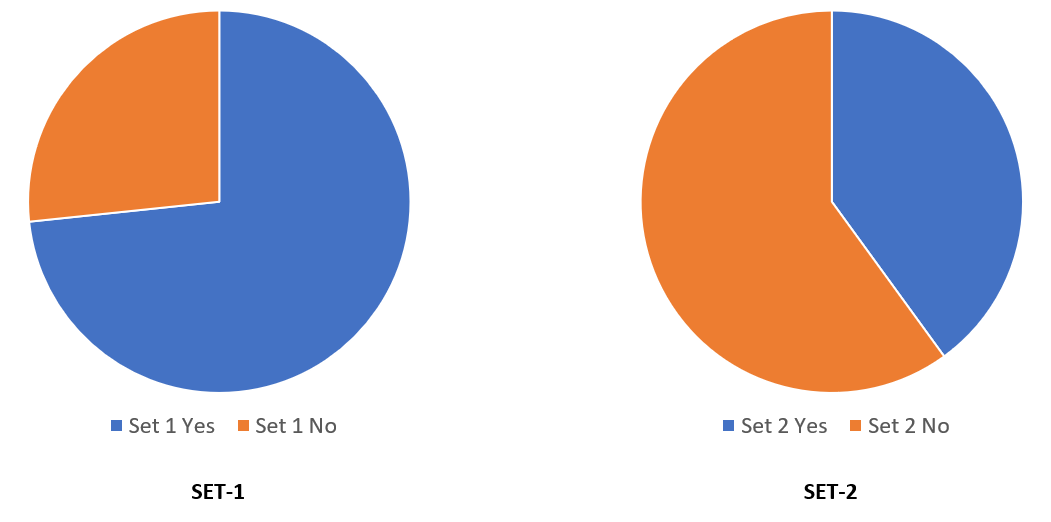}
    \caption{Responses to "Did you explored more about the topic after the session?"}
    \label{fig:mot}
\end{figure*}

Furthermore, the extent to which participants in Set-1 (VR-Based) and Set-2 (Traditional) explored the topic beyond what was covered in sessions was investigated to access motivation level instilled for learning. The data as depicted in Figure \ref{fig:mot} revealed a notable difference in the proportions of participants who engaged in additional exploration across the two sets.

In Set-1, a substantial 73.33\% of participants reported delving deeper into the topic after class, indicating a heightened curiosity and willingness to extend their learning independently. This contrasts with Set-2, where only 40\% of participants explored the topic further. The significant disparity in these proportions suggests that the immersive and interactive nature of the VR-based experience in Set-1 may have spurred a greater post-class engagement.

This finding aligns with the initial hypothesis that the dynamic and captivating elements of VR technology could enhance participants' motivation for self-directed learning. The ability to interact with celestial bodies in a virtual space may have stimulated a more profound interest, prompting participants to seek out additional resources and information. The higher proportion of exploration in Set-1 not only emphasizes the potential efficacy of VR technology in fostering post-class engagement but also underscores the importance of considering not just immediate learning outcomes but also the long-term impact on participants' interest and curiosity in the subject matter.

\begin{table}[htbp]
\centering
\begin{tabular}{c c c}
\hline
\textbf{Metric} & \textbf{Set-1 } & \textbf{Set-2 } \\
\hline
Engagement (\% reading more) & 80\% & 40\% \\

Average Class Experience Rating & 7.87 & 7.27 \\

Average Knowledge Improvement & 32.2 & 10.67 \\
\hline \\
\end{tabular}
\caption{Summary of Descriptive Analysis}
\label{tab:descanalysis}
\end{table}

\subsubsection{Engagement and Knowledge Improvement}

The engagement level of participants in both sets was assessed through their responses to the question, "Did you read more about the topic after class?" The majority of participants in Set-1 (VR-based experience) indicated that they read more about the topic after class, whereas in Set-2 (traditional teaching), the engagement in additional reading was comparatively lower. This suggests a potential positive impact of the VR-based experience on post-class engagement.

\subsubsection{Class Experience Ratings}

Participants in both sets provided ratings for their class experience on a scale of 10. These ratings serve as a subjective measure of the overall satisfaction and enjoyment of the learning experience. In Set-1, the ratings ranged from 6 to 9, with an average rating of 7.87. In Set-2, the ratings varied from 6 to 9, with an average rating of 7.27. While both sets generally received positive ratings, participants in Set-1 tended to rate their experience slightly higher.

\subsubsection{Knowledge Improvement Scores}

The knowledge improvement scores represent the difference between pre-test and post-test scores for each participant in both Set-1 and Set-2. In Set-1, improvements ranged from 13 to 47, with an average improvement of 32.2. In Set-2, improvements varied from 6 to 14, with an average improvement of 10.67. The higher average improvement in Set-1 suggests that the VR-based experience contributed more significantly to knowledge enhancement compared to the traditional teaching methods in Set-2.

\subsubsection{Comparative Analysis}

Comparing the results between Set-1 and Set-2, it is evident that the VR-based experience in Set-1 not only led to higher average knowledge improvement but also resulted in slightly more favorable class experience ratings. These descriptive findings lay the foundation for subsequent inferential statistical analyses, providing a preliminary understanding of the potential impact of the VSSE on participant engagement, class experience, and knowledge enhancement.

\subsection{Paired t-Test on knowledge improvement scores}
Results obtained from paired t-test are as given in Table \ref{tab:pairedttest}.

\begin{table}[htbp]
\centering
\begin{tabular}{lcc}
\hline
\textbf{Set} & \textbf{T-Statistic} & \textbf{P-Value} \\
\hline
Set-1 & $-11.87$ & $< 0.001$ \\
Set-2 & $-10.29$ & $< 0.001$ \\
\hline
\end{tabular}
\caption{T-Statistics for Paired Sample t-Tests}
\label{tab:pairedttest}
\end{table}

\subsubsection{Set-1}

For Set-1, the paired sample t-test revealed a statistically significant difference between the Pre-Test and Post-Test scores ($t = -11.87$, $p < 0.001$). The negative t-statistic indicates a notable decrease in scores from the initial assessment to the subsequent evaluation, providing compelling evidence against the null hypothesis.

\subsubsection{Set-2}

Similarly, in Set-2, the paired sample t-test demonstrated a significant difference between the Pre-Test and Post-Test scores ($t = -10.29$, $p < 0.001$). The negative t-statistic indicates a noteworthy decline in scores from the pre-assessment to the post-assessment, providing robust evidence against the null hypothesis.

\subsubsection{Null Hypothesis}

The null hypothesis ($H_0$) for each paired sample t-test asserts that there is no significant difference between the mean scores of the Pre-Test and Post-Test within each set. Mathematically, this is expressed as:
\[ H_0: \mu_{\text{Pre}} = \mu_{\text{Post}} \]

Where:
 \( \mu_{\text{Pre}} \) is the population mean of the Pre-Test scores and \( \mu_{\text{Post}} \) is the population mean of the Post-Test scores.

The rejection of the null hypothesis in both Set-1 and Set-2 suggests a substantial and statistically significant difference between the initial and subsequent assessments, providing evidence in favor of the effectiveness of the virtual reality based experiences.

\subsubsection{Interpretation}

These outcomes collectively suggest that, within both Set-1 and Set-2, there is a substantial and statistically significant decrease in scores following the educational interventions. The probability of observing such decreases due to random chance is exceedingly low ($p < 0.001$), supporting the contention that the interventions had a discernible impact on the participants' comprehension and knowledge.

\subsubsection{Implications for the Study}

The observed statistical significance underscores the efficacy of the educational interventions in fostering a tangible improvement in participants' understanding. These findings contribute valuable insights to the broader discourse on educational methodologies, emphasizing the discernible impact of both VR-based experiences and traditional methods on students' knowledge acquisition and retention.

\subsection{Correlation Analysis}
\begin{figure*}
    \centering
    \includegraphics[width=0.95\linewidth]{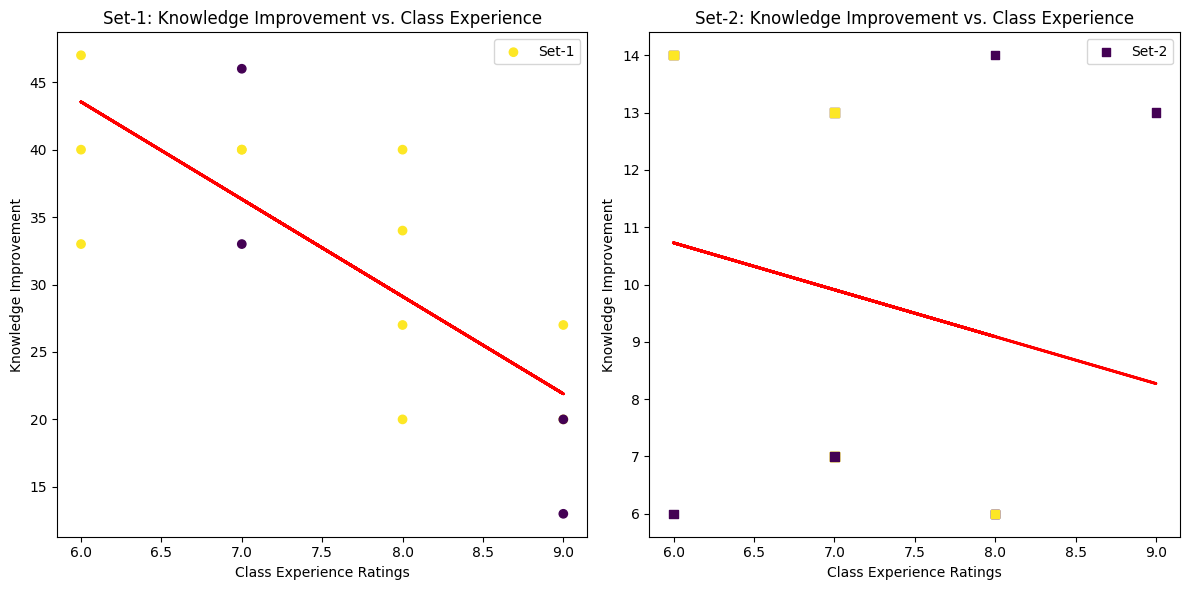}
    \caption{Analysis of relationship between class engagement levels and knowledge improvement levels}
    \label{fig:ce_ki}
\end{figure*}

Results obtained from correlation analysis are as given in Table \ref{tab:correlationanalysis}. Relationship between class engagement ratings and knowledge improvement levels have also been visualized in Figure \ref{fig:ce_ki}

\begin{table}[htbp]
    \centering
    \begin{tabular}{lcc}
        \hline
        \textbf{Correlation} & \textbf{Set-1} & \textbf{Set-2} \\
        \hline
        Knowledge Improvement vs. Class Experience & -0.77 & -0.20 \\
        Knowledge Improvement vs. Motivation & 0.24 & -0.02 \\
        \hline \\
    \end{tabular}
    \caption{Correlation coefficients between variables in Set-1 and Set-2.}
    \label{tab:correlationanalysis}
\end{table}

\subsubsection{Correlation between Knowledge Improvement and Class Experience (Set-1): -0.77}
A robust negative correlation ($r = -0.77$) between class experience ratings and knowledge improvement in Set-1 suggests a significant inverse relationship. Paradoxically, as class experience ratings increase, knowledge improvement tends to decrease. Potential contributing factors may include diverse learning styles or the need for tailored instructional strategies for optimal comprehension.

\subsubsection{Correlation between Knowledge Improvement and Motivation (Set-1): 0.24}
A modest positive correlation ($r = 0.24$) between motivation and knowledge improvement indicates a weak, yet discernible positive relationship in Set-1. Higher motivation levels are associated with slightly improved knowledge outcomes, underlining the nuanced interplay between motivation and academic performance.

\subsubsection{Correlation between Knowledge Improvement and Class Experience (Set-2): -0.20}
In Set-2, a marginal negative correlation ($r = -0.20$) between class experience ratings and knowledge improvement is observed. However, the correlation is weak, suggesting that class experience ratings may not serve as a robust predictor of knowledge improvement in this context.

\subsubsection{Correlation between Knowledge Improvement and Motivation (Set-2): -0.02}
The minimal and near-zero correlation ($r = -0.02$) between motivation and knowledge improvement in Set-2 suggests a lack of discernible linear relationship. Motivation levels do not exhibit a clear impact on knowledge improvement, highlighting potential complexities in the learning dynamics of this set.

\subsection{Impact of Class Experience Ratings on Knowledge Improvement}
ANOVA analysis revealed that class experience ratings significantly influence knowledge improvement in both traditional and virtual educational settings. The findings emphasize the potential of VR-based educational experiences, as exemplified in Set-1, to positively shape class experiences and enhance educational outcomes. The nuanced distinctions observed between the two sets offer avenues for future research to unravel the multifaceted nature of learning dynamics in diverse instructional contexts.

\subsubsection{Set-1}
The analysis of Set-1, where participants engaged with a Virtual Solar System Experience (VSSE), reveals a highly significant relationship between class experience ratings and knowledge improvement. The obtained ANOVA p-value ($p < 0.001$, $9.34 \times 10^{-10}$) underscores the statistical significance of the observed differences in knowledge enhancement based on varying class experience evaluations.The remarkably low p-value signifies that the quality of class experience, as perceived by participants, exerts a substantial influence on the degree of knowledge improvement in the context of the VSSE. This implies that positive class experiences, potentially shaped by the virtual learning environment, contribute significantly to the efficacy of educational outcomes.

\subsubsection{Set-2}

In Set-2, where a comparative analysis between traditional teaching methods and the VSSE was conducted, the ANOVA results remain statistically significant, albeit with a moderately higher p-value ($p = 0.0111$). This indicates a meaningful but relatively less pronounced impact of class experience ratings on knowledge improvement compared to Set-1. The elevated p-value in Set-2 suggests that while class experience ratings still play a significant role in knowledge enhancement, other factors inherent to traditional teaching methods may contribute to the observed effects. The dynamics of knowledge improvement in this set appear to be influenced by a combination of instructional approaches.

\subsubsection{Comparative Analysis}

\begin{enumerate}
    \item \textbf{Magnitude of Impact:} Set-1 exhibits a stronger association between class experience ratings and knowledge improvement, highlighting the effectiveness of the VR-based educational approach.
    \item \textbf{Practical Significance:} In both sets, the statistically significant findings prompt considerations of practical significance. While Set-1 demonstrates a robust impact, the somewhat attenuated effect in Set-2 may indicate nuances in the educational dynamics.
    \item \textbf{Implications for Educational Design:} These results underscore the importance of considering the design of educational experiences, particularly in virtual environments, to optimize class experiences and subsequently enhance knowledge acquisition.
    \item \textbf{Further Investigation:} The observed differences between Set-1 and Set-2 invite further exploration into the specific pedagogical components influencing knowledge improvement. Qualitative assessments and detailed analysis of teaching methodologies can provide nuanced insights.
\end{enumerate}

\subsection{Regression Analysis}
\subsubsection{Set-1}

The linear regression analysis for Set-1, where participants exclusively engaged in the Virtual Solar System Experience (VSSE), produced a p-value of 0.0006907. This p-value indicates a statistically significant linear relationship between class experience ratings and knowledge improvement. The low p-value in Set-1 underscores a robust linear association between positive class experiences, as perceived within the VSSE, and notable knowledge improvement. This highlights the crucial role of immersive virtual experiences in positively influencing educational outcomes.

\subsubsection{Set-2}

In Set-2, which involved a comparative analysis between traditional teaching methods and the VSSE, the linear regression analysis resulted in a p-value of 0.4678, suggesting a lack of statistically significant linear relationship between class experience ratings and knowledge improvement. The higher p-value in Set-2 implies that, in the presence of both traditional and virtual teaching methods, the linear impact of class experience ratings on knowledge improvement is not statistically discernible. This suggests the presence of complex dynamics or potential non-linear relationships influencing knowledge gains.

\subsection{User Feedback on the Usability of Virtual Reality (VR) Experience}
In assessing the usability of the Virtual Solar System Experience (VSSE), feedback from participants provides valuable insights into the user experience. User feedback not only provides valuable insights into the usability of the VR system but also aligns with and complements the quantitative findings. The positive sentiments expressed by participants resonate with the statistically significant impact observed in Set-1, underscoring the potential of user-friendly and engaging VR experiences in educational contexts. The following structured summaries encapsulate the sentiments expressed by students, shedding light on the usability aspects of the VR interface.

\subsection{Intuitiveness of Controls}

A recurring theme among participants is the intuitiveness of the VR controls. Users consistently praised the ease with which they navigated through the solar system, emphasizing the user-friendly nature of the interface. However, a small subset of participants expressed initial confusion with certain controls, highlighting the need for clearer instructions. This aligns with the findings of the linear regression analysis in Set-1, where a significant linear relationship between positive class experiences and knowledge improvement was observed. The intuitiveness of controls potentially contributes to a more engaging and effective learning environment.

\subsection{Visual Appeal and Engagement}

Feedback consistently highlights the visual appeal of the VR system, with students expressing that the visuals were stunning and made learning about the solar system more enjoyable. However, a few participants found certain visual elements distracting, suggesting that optimizing visual clarity could enhance the overall experience. This aligns with the positive impact observed in Set-1, where participants exclusively engaged in the VR experience. The statistically significant linear relationship between positive class experiences and knowledge improvement may be attributed, in part, to the engaging and visually appealing nature of the VR interface.

\subsection{Usability for Novice Users}

Noteworthy is the positive feedback from participants who were initially unfamiliar with VR technology. Comments indicate that the VR system was surprisingly easy to use, even for those new to virtual reality. However, a subset of participants struggled with initial setup procedures, emphasizing the importance of comprehensive user guides. This is consistent with the findings in Set-2, where the linear relationship between class experience ratings and knowledge improvement was not statistically significant. The usability for novice users may be a critical factor, especially in mixed-method educational settings.

\subsection{Interactive Features and Self-Paced Learning}

Users appreciated the interactive features that allowed them to explore different aspects of the solar system at their own pace. This resonates with the qualitative insights obtained from student feedback, emphasizing the user-friendly approach to education through self-paced exploration. Some participants suggested additional interactive elements to deepen the learning experience. Such features likely contribute to the positive holistic impact observed in the comparative analysis.

\section{Limitations}
This study has certain limitations that should be considered. The findings may be more applicable to specific contexts, like computer science education, and might not easily extend to other educational settings. Additionally, the rapid evolution of VR technology means that our results could be influenced by future changes in this technology. While the use of VR in education has shown a clear surge in user satisfaction compared to conventional teaching methods, there remains a critical gap in demonstrating a measurable increase in learning rates through VR applications \cite{p24}. The challenge lies in establishing VR as a pedagogically robust tool that not only captivates learners but also significantly enhances learning outcomes. The pivotal challenge facing educators and researchers is to establish VR as more than a novel and engaging technology; it must be validated as a pedagogically robust tool with demonstrable efficacy in fostering meaningful learning.

The effects of the perceived quality of the virtual environment's graphics are also crucial to the impact of immersive environments on the participant \cite{p27}. While the satisfaction of users attests to the appeal of VR, the next crucial step is to bridge the gap between engagement and measurable learning gains. This necessitates a comprehensive exploration of the ways in which VR can be strategically integrated into educational practices to elicit quantifiable advancements in students' understanding, retention, and application of knowledge.

Furthermore, focusing on a VR-based solar system experience may limit the broad application of our conclusions to other educational uses of VR. It's crucial to note that individual learner differences weren't extensively explored within the scope of this research. Despite these limitations, our study provides valuable insights into the intersection of VR and education. It lays the groundwork for future research to delve deeper into how immersive technologies impact various aspects of learning.

\section{Conclusion and Future Scope}
This paper presents a rigorous examination of the impact of Virtual Reality (VR) on educational outcomes, focusing on a VR-based solar system experience. Through a comprehensive mixed-methods approach, incorporating both qualitative and quantitative analyses, this study provides empirical evidence supporting the efficacy of VR in enhancing critical aspects of the learning process. The results demonstrate that participants engaged in the VR-based experience exhibited a significant increase in knowledge improvement compared to traditional teaching methods. The observed negative correlation between class experience ratings and knowledge improvement in the VR-based set underscores the importance of carefully designed immersive experiences, as perceived class satisfaction positively influences educational outcomes. Additionally, the study identifies a notable difference in post-class engagement, with a higher proportion of participants in the VR-based set exploring the topic further. This highlights the enduring impact of immersive VR experiences on fostering sustained interest and curiosity beyond the immediate learning environment. The comparative analysis between the VR-based set and traditional teaching methods reveals the unique strengths of VR in positively influencing class experiences and subsequent knowledge improvement. However, it also indicates the need for further exploration into the nuanced dynamics of learning outcomes influenced by different methodologies.

Future research avenues include an in-depth exploration of the pedagogical components inherent in VR experiences that contribute to enhanced knowledge improvement. Qualitative assessments and detailed analyses of instructional strategies can provide valuable insights into the mechanisms driving the observed effects. Moreover, investigating the transferability of findings to different educational contexts and subjects will contribute to a more comprehensive understanding of VR's potential as a transformative tool in education.

\section*{Acknowledgment}We extend our sincere gratitude to Mr. Kushal Jain, IIIT Naya Raipur for his invaluable assistance in providing essential hardware support and thoughtful feedback.
% -------------------------------INSERT REFERENCES HERE----------------------------

\end{document}